\documentclass{article} 
\usepackage[dvips]{graphicx}

\def\ra{\rightarrow}
\def\be{\begin{equation}} 
\def\ee{\end{equation}}

\begin{document}

\begin{center}
\bf{ Gluonium states and the Pomeron trajectory}\\
\vspace{3mm}
\rm{ M. N. Sergeenko }\\
\vspace{2mm}
\it{Institute of Physics, The National Academy of Sciences
of Belarus\\ 70 F. Skaryna Avenue, Minsk, 220072, Belarus, and\\
State University of Transport,\\ 
34 Kirova Street, Gomel, 246653, Belarus}\footnote{
The present address of the author.\ E-mail: msergeen@gmail.com}
\end{center}

\begin{abstract}
Pomeron is modeled as a system of two interacting by Cornell 
potential massive gluons. In bound state region, relativistic 
wave equation for the potential is analized. Two exact asymptotic 
solutions of the equation are used to derive an interpolating 
mass formula for gluonium states and the Pomeron trajectory in 
the whole region. The trajectory obtained is linear at large 
timelike $t$ and flattens off at $-1$ in the scattering 
region at large $-t$. Parameters of the trajectory are found 
from the fit of recent HERA data for $\alpha_P(t)$.\\ 

\noindent PACS number(s): 11.15.Me, 11.15.Pg, 11.15.Tk \\

\end{abstract}

\section{ Introduction}	

As known many features of hadron interactions at high energies are
remarkably well reproduced by Regge calculus and, in particular, the
high-energy behavior of hadron scattering is very well predicted by
the Pomeron \cite{DLPW}. The well known ``soft'' Pomeron is responsible 
for most of the total cross section in hadron-hadron collisions, for 
small $-t$ elastic scattering and for diffractive dissociation. Hadron
amplitude at high energy is proportional to $s^{\alpha_P(t)}$, where
$\alpha_P(t)$ is the Pomeron trajectory.

Last years the phenomenology of processes dominated by the perturbative 
Regge kinematics attracts increasing interest \cite{Kir}. The most 
prominent example is the so-called BFKL-Pomeron (``hard'' Pomeron) 
\cite{Fa,Lip} which has been found very useful in describing the rise of 
nucleon structure function $F_2(x,Q^2)$ at small $x$ \cite{Wsh}, gives
excellent description of data for $J/\Psi$ photoproduction and the
charm structure function $F_2^c$ \cite{DL}. It is expected also to
show up in jet-inclusive final states with rapidity gaps in
deep-inelastic, hadron-hadron interactions and other processes
\cite{Mu}.

Most important parameters of the Pomeron in high-energy 
hadron physics are the intercept $\alpha_P(0)$ and the slope 
$\alpha^\prime_P(0)$ of the Pomeron trajectory. Usually, the parameters 
are determined from experiment \cite{Pexp,CollGa}. However, the ideas 
proposed in Ref. \cite{DL185} were rather successful. They have 
succeeded, at $t=0$, to describe the Pomeron in terms of modified 
tree-level two-gluon exchange, thus providing a connection with QCD. 
This approach has shown, that the Pomeron trajectory could be obtained 
at non zero $t$, when resumed to higher orders, whereas perturbative 
QCD (pQCD) seems to fail in that respect.

Most recent small $-t$ ZEUS data for exclusive $\rho$ and $\phi$
photoproduction \cite{ZEUS,Ast} lead to a slope $\alpha^\prime_P(0)$
for the trajectory of the ``soft'' Pomeron that differs significantly
from the classical value $\alpha^\prime_P(0)=0.25$\,GeV$^{-2}$. The
results of these experiments have been discussed in Ref. \cite{DLZ}.
It was pointed out that the slope $\alpha^\prime_P(0)$ of the ``soft''
Pomeron should be determined from the data at small $-t$,
$-t\simeq 0.4$\,GeV$^{-2}$ at HERA energy \cite{HERA}, where ``soft''-Pomeron
exchange dominates the differential cross section; but recent ZEUS
measurements \cite{ZEUS} extend to rather large $-t$. The ZEUS data
have been explained in Ref. \cite{DL} by adding in a flavor-blind
``hard''-Pomeron contribution, whose magnitude is calculated from
the data for exclusive $J/\Psi$ photoproduction. These data can be
explained by nonlinear behavior of the Pomeron trajectory we obtain
in this work.

In this work, we obtain the Pomeron trajectory in the whole region.
We use the method suggested earlier to calculate the Reggeon 
trajectories \cite{SeA,SeZ}. The Pomeron is considered as the 
$t$-channel process. Then we take a picture 
where Pomeron is dual to glueball (gluonium) states, i.e. it is 
a bound state in s-channel. We work in the framework of the potential 
approach, which is a {\em natural framework} for the studying Regge 
trajectories and their properties. 

In our model, two massive gluons interact by means of the QCD 
motivated funnel type potential \cite{Sim}. 
In the scattering region ($t\le 0$), this system corresponds to a
ladder-type diagrams, i.e. tree-level two-gluon exchange. In bound 
state region ($t>0$) we have a bound state problem for two
massive gluons, i.e., we consider gluonium and its excited states.

Gluonium masses are calculated with the use of the interpolating 
mass formula we obtain from two exact asymptotic solutions 
of relativistic wave equation \cite{SeZ}. Inverting this formula, 
we derive an analytic expression for the Pomeron trajectory,
$\alpha_P(t)$. In the scattering region, at $-t\gg\Lambda_{QCD}$,
the trajectory flattens off at $-1$, i.e., it has asymptote
$\alpha_P(t\ra -\infty)=-1$. In bound state region, at large 
timelike $t$, the Pomeron trajectory is linear in accordance with 
the string model.

\section{The ``soft'' and ``hard'' Pomerons}

There exist different approaches to investigating an object such
as the Pomeron; the issue of ``soft'' and ``hard'' Pomeron has
been discussed extensively in the literature 
\cite{Fa,Lip,DuKaSi}. The classic ``soft'' Pomeron is
constructed from multi-peripheral hadronic exchanges and has
intercept $\alpha_P(0)=1$. Because this is not compatible with the
rising hadronic cross sections at high energies, the ``soft''
Pomeron was replaced by a ``soft'' supercritical Pomeron with an
intercept $\alpha_P(0)>1$, i.e. $\alpha_P(0)-1=\Delta >0$
\cite{Land,Kaid}.

In the framework of QCD (in the scattering region), the Pomeron is
understood as the exchange of two (or more) gluons \cite{LowN}. The
perturbative QCD approach to the Pomeron has been discussed in
Refs. \cite{Fa,Lip}. The Balitski\v{i}-Fadin-Kuraev-Lipatov (BFKL)
Pomeron \cite{Fa} is built out of multiperipheral high transverse
momentum gluon exchanges and predicts a different Pomeron, which in
complex $j$ plane has a series of poles at $1<j<1+\Delta$ with
$\Delta\simeq 0.45$. This approach gives good results where the
leading logarithmic approximation (LLA) holds.

The leading and subleading logarithmic summation in QCD \cite{Lip}
predicts asymptote $\alpha_P(t)=1$ for the ``hard'' BFKL-Pomeron at 
large spacelike $t$ \cite{Fa,Lip},
\be
\alpha_P(t)\simeq 1 + O(\tilde g^2(t)), ~~~t\ra -\infty,
\label{QCDas} 
\ee
where $\tilde g^2(t)$ stands for the running  coupling of QCD. 

This prediction for the Pomeron trajectory contradicts to present 
ISR and Tevatron data, and most recent ZEUS data \cite{ZEUS,Ast} on
$\alpha_P(t)$. The shrinkage of the forward elastic differential
cross section peak in $pp/p\bar p$ elastic scattering from ISR to
Tevatron energies means that the Pomeron trajectory is approximately
linear with a slope $\propto 0.25$\,GeV$^{-2}$ out to $t\simeq
-1.0$\,GeV$^2$. The ISR diffractive dissociation data indicate that
this approximate linearity continues at least out to $t\simeq
-2.0$\,GeV$^2$ by which time the value of the trajectory is $\propto
0.6$, i.e. considerably below $1$. Recent ZEUS data \cite{ZEUS}
demonstrate similar behavior.

In the last years difficulties have emerged with the BFKL equation
(see \cite{DS} and references therein). The LLA BFKL predictions
overestimate the $\gamma^*\gamma^*$ cross section by a large factor
\cite{DS}. In the BFKL formalism, there is a problem at LL order in
setting the two mass scales on which the cross section depends: the
mass $\mu^2$ at which the strong coupling $\alpha_s$ is evaluated and
the mass $Q_s^2$ which provides the scale for high-energy logarithms;
the results are very sensitive to these parameters \cite{BHS}. An
additional uncertainty is due to the correct treatment of the
production of massive charm quarks. An attempt to overcome the scale
problem reduces both the size of the BFKL cross section and its energy
dependence \cite{BBPR}.

A comparative investigation of various Pomeron models was carried out
in the impact parameter space through their predicted values of 
$\sigma_{tot}$, slope $B$, and $\sigma_{el}/\sigma_{tot}$ in high 
energy $pp$ and $p\bar p$ scattering \cite{Lev}. The main result of this 
investigation is that the data is
compatible with a smooth transition from a ``soft'' to a ``hard'' 
Pomeron contribution which can account for the rise of $\sigma_{tot}$ 
with $s$. Note, however, that the singularity of the ``hard'' Pomeron 
is not a pole but a cut in the $\alpha_P(t)$ plane, starting at
$\Delta=12\alpha_s{\rm ln}2/\pi$ \cite{Fa,Bjor}. Then, if ``soft'' and
BFKL Pomeron have a common origin, the discontinuity across the cut
in the $\alpha_P(t)$ plane must have a strong $t$ dependence
\cite{Bjor} that points out on nonlinearity of the Pomeron trajectory.

Both the ``soft'' supercritical and the ``hard'' QCD Pomeron predict 
a powerlike rise of the total cross section, $\sigma_{tot}\propto
s^\Delta$. However, the physics of the ``soft'' Pomeron is much less
clear; any realistic pQCD attempts lead to the introduction of an
``infrared'' cutoff in gluon propagator that generates the main part
of Pomeron exchange. The derivation of nonperturbative (NP) gluon
propagator from QCD \cite{Land,Halz,Kot} contains many assumptions
to be proved. In many of these studies, the low-momentum singularity
of the gauge field propagator is softened. A soft infrared behavior
can be also obtained in a Yang-Mills theory with Higgs mechanism,
where the gluon acquires a mass, or for some solutions of the
Dyson-Schwinger equation \cite{Halz}, or simply in the presence of
a cutoff.

There are currently no any theoretical estimations of the Pomeron 
trajectory. The behavior of the Pomeron trajectory $\alpha_P(t)$ 
in the whole region is unknown. The linear dependence,
\be
\alpha_P(t)=\alpha_P(0)+\alpha^\prime_P(0)t,  \label{Ptlin}
\ee
is usually assumed, which is a good approximation in small $-t$
region. It is a goal of this work to reproduce the Pomeron
trajectory in the whole region and calculate the parameters
$\alpha_P(0)$ and $\alpha^\prime_P(0)$.

There are good bases to believe that the ``soft'' Pomeron can be
considered as an exchange of two NP gluons, whose properties are
dictated by the expected structure of the QCD vacuum \cite{Land}.
In recent time several successful models based on the theory of
the ``soft'' supercritical Pomeron have been developed in which
the Pomeron is modeled as an exchange of two NP gluons dynamicaly 
generated mass whose propagator is finite at $q^2=0$ \cite{Halz,LySe}. 
These models and their modification \cite{SeNP} provide a reasonable 
description of data in and above CERN Intersecting Storage Rings 
(ISR) energy range.

\section{Some general properties of Regge trajectories}

Let us remind some general results obtained for the Regge
trajectories in recent time. This will allow us to understand many
features of such a Reggeon known as the Pomeron.

It is well-known experimental fact that hadrons populate linear Regge
trajectories in $s$-chanel (or $t>0$). That means that the square of 
the mass of a state with orbital angular momentum $l$ is proportional 
to $l$: $M^2(l)=\beta l+const$, with the same slope, 
$\beta\simeq 1.2$\,GeV$^2$, for all trajectories. There exists a 
conviction, that the Regge  trajectories $\alpha(t)$ are linear in 
the whole region, that is, not only in the bound state region ($t>0$) 
but in the scattering region ($t<0$), too. 

Main characteristics of ``soft'' hadronic processes in the scattering 
region can be understood in terms of the exchange of particles, which 
lie on linear Regge trajectories. This approximate linearity 
encouraged the dual model approach to strong interactions which in 
tree approximation assumes exactly linear trajectories. However, 
the conception of linear Regge trajectories is not consistent with 
experimental data and expectations of pQCD at large spacelike momentum 
transfer $-t\gg\Lambda_{QCD}$. In the experiment \cite{Bar} far more
complicated behavior of the $\rho$ meson trajectory,
$\alpha_\rho(t)$, was discovered; the $\rho$ trajectory flattens off
at about $-0.6$.

There are different approaches to investigate the Regge trajectories
\cite{Kir,Fa,Lip}. Significant efforts were undertaken in order 
to obtain information on the behavior of Regge trajectories at 
large spacelike $t$ \cite{Lip,CDL,CoKe,BaSh}; the asymptotic behavior 
at $-t\ra\infty$ has been discussed by many authors
\cite{Ans,Pet,Bro,SeI}. As shown in \cite{CoKe},
exclusive  processes whose cross-sections are determined by Regge
pole trajectory exchange, $\alpha(t)$, at small momentum transfers,
$t$, are controlled by these same exchanges at very large spacelike
$t$, too. Hereby trajectory must to be nonlinear and one of the most
crucial distinction between small $-t$ behavior of $\alpha(t)$ and
large $-t$ behavior of $\alpha(t)$ involves the asymptotic form of
Regge trajectories at $-t\ra\infty$.

Important information on large $-t$ behavior of trajectories can be
obtained from the comparing the predictions for the scattering
amplitude, $T(s,t)$, of the "quark counting rule" \cite{Mat} and the
Regge pole approach at $s\ra\infty$, $-t$ fixed. From the
demand of a smooth interpolation between these two predictions the
condition was obtained \cite{Pet},
\be
\alpha(t) = const, ~~~ t \ra -\infty. \label{atcons}
\ee
Such asymptotic behavior of the Regge trajectories was proposed by
various authors \cite{CoKe,Pet,Bro,Mat}, and seems does not contradict
to experimental data \cite{Bar}.

There have been undertook considerable efforts to extend the
constituent interchange model (CIM) (see Ref. \cite{Siv}) from the
fixed-angle region into the fixed $t$-region. These efforts have
resulted in the prediction for the large $-t$ behavior of $\rho$
trajectory
\be
\alpha_{\rho}(t)=-1, ~~~t\ra -\infty. \label{at1}
\ee
The same asymptotic behavior (\ref{at1}) for all leading $S=1$
quarkonium Regge trajectories was obtained in our Ref. \cite{SeA} on 
the basis of analysis of the relativistic quasipotential equation with
the QCD motivated potential, and in Ref. \cite{SeZ} on the basis of
solution of the Klein-Gordon equation. The Regge trajectories have been
determined as the function $l(E^2)$, where $E^2$ is the squared
quarkonium mass obtained from solution of the eigenvalue problem for
two bound quarks and $l$ is the relative orbital angular momentum of
the quarks. Main result of these investigations is that, that the Regge 
trajectories are nonlinear and have the asymptote (\ref{at1}).

The Pomeron trajectory can be obtained similar way as the Reggeon ones.
However, instead of two interacting quarks we consider two interacting 
massive gluons. We deal with the gluon system primarily in bound state 
region and solve relativistic wave equation for the funnel-type QCD
motivated potential. Then, using the same technic as for Regeons, we
obtain the Pomeron trajectory $\alpha_P(t)$ in the whole region 
$-\infty <t<\infty$.

\section{The model}

Let us outline the main features of our model. Two interacting 
particles can be considered both in the scattering region ($t\le 0$) 
and in bound state region ($t>0$). The wave equation describes 
these two interacting particles in the whole region of $t$, 
$-\infty<t<\infty$, with the corresponding boundary conditions for 
the wave function. Most important aspect in this approach is the
form of the potential. 

The potential approach is a more convenient and, at the same time, 
simpler way to reconstruct the Regge trajectories. Moreover, the Regge 
theory itself was proposed by Regge at the dealing with solutions of 
the Schr\"odinger equation for the {\em nonrelativistic} potential 
scattering. In other words, the potential approach and studying 
solutions of the wave equation is a natural framework to investigate 
Regge trajectories and its properties, in spite of the phenomenological 
nonrelativistic nature of the potential.

Because of the intrinsically nonperturbative nature of bound-state
problem in non-Abelian gauge theories, it is up to now, not possible
to derive the forces acting between the quarks and gluons from first
principles. It has been well tested that hard processes are governed 
by short range part of the strong interaction. It is generally agreed 
that, in pQCD, as in QED the essential interaction at small
distances is instantaneous Coulomb exchange; in QCD, it is $qq$,
$qg$, or $gg$ Coulomb scattering \cite{Bjor}. The dynamics is the
Coulomb interaction,
\be
V_S(r)=-\frac{\tilde{\alpha}}r, ~~~ r\ra 0,  \label{Coul}
\ee
where $\tilde\alpha$ is the effective strong coupling constant
($\tilde\alpha=\frac 43\alpha_s$ for mesons). In momentum space,
the Coulomb potential (\ref{Coul}) is $\tilde V(t)=4\pi\tilde\alpha/t$, 
that is the Fourier transform of $\tilde V(t)$ leads to the potential 
(\ref{Coul}). At large $-t=-q^2$, the potential (\ref{Coul}) corresponds 
to the scattering amplitude in the Born approximation with the gluon 
propagator $D(q^2)\propto 1/q^2$ \cite{Bjor}. 

For large distances, in order to be able to  describe  confinement,
the potential has to rise to infinity. From lattice-gauge-theory
computations \cite{BaSh} follows that this rise is an approximately 
linear, that is,
\be 
V_L(r)\simeq\kappa r,~~~ r\ra  
\infty, \label{Conf}
\ee
where $\kappa\simeq 0.15\,GeV^2$ being the string tension.

The potential is more poorly understood in an intermediate region. 
In this region, many well known potentials give reasonable results 
for hadron masses (see \cite{Luch} and Refs. therein), but these 
results do not depend very strong on the form of the potential. 
The most reasonable possibility to construct an interquark potential, 
which satisfies both of the above constraints, is to simply add these 
two contributions. This leads to the so-called funnel-shaped (or
Cornell) potential \cite{Eich}:
\be
V(r)=-\frac{\tilde{\alpha}}r +\kappa r +V_0. \label{Corn}
\ee
A closer inspections reveals that all phenomenologically acceptable
"QCD-inspired" potentials are only variations around the funnel
potential. Its parameters are directly related to basic
physical quantities: the universal Regge slope $\alpha'\simeq
0.9\,($GeV$/c)^{-2}$ for trajectories and one-gluon-exchange
coupling strength $\alpha_s$ at small distances. As for constant
$V_0$, usually, it is added to the confining contribution. 

It is usually supposed a distinctly different dynamical origin 
of Pomeron and Reggeons. This suggests a different dependence 
$\alpha_P(t)$ and the ordinary Reggeons. The physics of
ordinary Reggeon (such as the $\rho)$ is exchange of a ladder, for
which the sides are generally regarded to be constituent quarks. The
rungs of the ladder represent the binding potential between quarks
(nonperturbative gluons), i.e. Reggeons can be obtained as orbital
(and radial) excitations of bound system of interacting quarks
\cite{Bjor}. Similarly, due to the nonabelian nature of gluonic
field, gluons interact each other. This (and strong believe that
gluon has nonzero mass) makes it possible to consider the bound 
state problem for two (or more) gluons interacting by means of QCD
motivated potential, i.e. gluonium excited states (glueballs).

Glueballs were suggested theoretically in Refs. \cite{Min} and then
have been extensively studied in the framework of different
approaches (see \cite{Di}), and potential-like models \cite{Rob},
too, where a notation of the constituent gluon mass $\mu_0$ was
introduced. There is no overall agreement between the results of
different approaches in the theory (and also in experiment);
moreover, many theoretical approaches contain basic model assumptions
which are difficult to prove starting from the QCD Lagrangian.

Gluonium masses have been calculated in \cite{Sim} within a new
method called the Vacuum Correlator Model (VCM) \cite{Si}. In this
model, all nonperturbative and perturbative dynamics of quarks and
gluons in the Pomeron is universally described by lowest cumulants,
i.e. gauge invariant correlators of the type $\langle F_{\mu\nu}(x_1)
\ldots F_{\lambda\sigma}(x_{\nu})\rangle$. Linear confinement
naturally arises due to the presence of specific structures in the
correlators, which are nonzero for Abelian fields when monopole
condensate is formed, and for nonabelian fields due to their
specifics nonabelian structure. 

The gluon correlators (cumulants) are characterized by a specific 
decay time (correlation time), which is a universal feature of 
the QCD vacuum \cite{Si}. For quarks and gluons the situation is 
{\em the same}. All the methods and formulae derived for
light quarks apply also for gluons some self-evident replacements 
\cite{Sim}. The Regge trajectories for mesons and baryons have been 
found with good phenomenological properties, i.e. the slope is the 
same for mesons and baryons and is equal to $1/8\sigma$, where $\sigma$ 
is the string tension \cite{SiL}. In this way one connects the properties 
of the QCD Pomeron with the Regge phenomenology.

The physical picture of gluonium can be formulated as follows. In the
vacuum, two or three gluons are excited. Each of gluons propagates
through the Pomeron feeling both perturbative and nonperturbative
interactions with this medium and another gluon. This gives rise to
the Coulomb-like and string-type interactions, which are handled as
in the case of light quarks \cite{SiL}.

In this model all dependence on Pomeron gluonic fields $\bar A_\mu$ is
contained in the adjoined Wilson loop $\langle W_{adj}(C)\rangle$,
where the closed contour $C$ runs over trajectories $z_\mu(\sigma)$
and $\bar z_\mu(\sigma)$ of both gluons. For the nonperturbative part
and Coulomb interactions the ratio of correlators at least at small 
distance in adjoined and fundamental representation is the number
$N_2=C_2^{adj}/C_2^{fund}$, where $C_2^{adj}=N_c=3$,
$C_2^{fund}=(N^2_c-1)/2N_c$. In the adjoined and fundamental 
representation, the final form of interaction of two gluons is given 
by \cite{Sim}
\be
V^{adj}(r) =-\frac{\alpha_a}r +\sigma_a r -C_0, \label{Vadj}
\ee 
where $\alpha_a \equiv\alpha^{adj}=\frac 94\frac 43\alpha_s^{fund}$
$\equiv 3\alpha_s^{fund}$, $\sigma_a \equiv\sigma^{adj}$= $\frac
94\sigma^{fund}$; $\alpha_s^{fund}$ is the strong coupling,
$\sigma^{fund}\equiv\sigma\simeq 0.15$\,GeV$^2$ is the string
tension, and $C_0$ is the arbitrary parameter. We see that the slope
of the gluonium trajectory at large $l$ is $\frac 49$ of that of
light meson trajectories. Important feature is the occurrence of the
constituent gluon mass $\mu_0$ in the dynamical equation, which is
state dependent. For lowest state ($l=0$, $n_r=0$) this model gives
$\mu_0=\frac 32 m_q$, where $m_q$ is the quark mass.

In hadron physics, the nature of the potential is very important.
Concept of the scalarlike potential is especially important in
hadron physics. In relativistic potential models of quarkonia based
on a Dirac-type equation with a local potential is a sharp
distinction between a linear potential $V(r)$ which is vectorlike and
one which is scalarlike. There are normalizable solutions for a
scalarlike $V(r)$ but not for a vectorlike $V(r)$ \cite{Su,SC}. No any
problems arise and no any difficulties encountered with the numerical
solution if the confining potential is purely {\em scalarlike}. It
was shown in many works that the effective interaction has to be
scalar in order to confine particles inside the hadrons (see, for
example, Refs. \cite{SC}). 

The same is true for the short-range Coulomb potential. In Ref. 
\cite{SePA}, we have considered nonrelativistic semiclassical 
wave equation for central potentials, and in Ref. \cite{SeM} - solution 
of relativistic one for vectorlike and scalarlike Coulomb potentials. 
we have obtained two results for the Coulomb potential: the known 
exact result for spinless particles which coincides with one obtained
from solution of the Klein-Gordon equation and another result
obtained from solution of the semiclassical equation for the
scalarlike Coulomb potential. We have shown that, unlike the known
relativistic wave equations for the Coulomb potential, the
semiclassical equation with the scalarlike potential has the regular
solution at the spatial origin \cite{SeM}. Thus, in this work, for 
the $gg$ interaction, we use the scalarlike Cornell potential 
(\ref{Vadj}).

We use the potential (\ref{Vadj}) to reproduce the gluonium masses. 
For $gg$ system, to be able reproduce the Pomeron trajectory, we need 
to obtain an analytic expression for the squared gluonium mass 
$E^2$. For this, we solve the relatvistic semiclassical wave 
equation \cite{SePA,SeM}; for two interacting gluons of equal masses 
$\mu_1=\mu_2=\mu_0$, the semiclassical wave equation with the scalarlike 
potential (\ref{Vadj}) is
\be 
(-i)^2\left(\frac{\partial^2}{\partial r^2} +\frac 1{r^2}
\frac{\partial^2}{\partial\theta^2} +\frac 1{r^2\sin^2\theta}
\frac{\partial^2}{\partial\varphi^2}\right)\tilde\psi(\vec r) = 
\left[\frac{E^2}4 
-[\mu_0+V^{adj}(r)]^2\right]\tilde\psi(\vec r). \label{eq}
\ee
It is hard to find the analytic solution of equation (\ref{eq}) for
the potential (\ref{Vadj}). But we can find exact analytic solutions 
of this equation for two asymptotic limits of the potential 
(\ref{Vadj}), the Coulomb and the linear potentials \cite{SeM}:
\be
E_n^2 =4\mu_0^2\left[1 -\frac{\alpha_a^2}{\left(n_r+\frac 12 +
\sqrt{(l+\frac 12)^2 +\alpha_a^2}\right)^2}\right], \label{ECoul}
\ee
\be
E_n^2 =8\sigma_a\left(2n_r+l -\alpha_a +\frac 32\right).
\label{Elin} 
\ee
At small distances, where the Coulomb type contribution dominates,
the effective strong coupling, $\tilde\alpha$, is a small value and
Eq. (\ref{ECoul}) can be written in the simpler form
\be 
E_n^2 \simeq \mu_0^2\left[1 -\frac{\alpha_a^2}{(n_r+l+1)^2}\right]. 
\label{ECoS}
\ee
Note, that equation (\ref{ECoS}) for the squared invariant mass of
the system of two particles has the correct relativistic form, i.e.
$E_n^2 =4(p_n^2+\mu_0^2)$, where $p_n^2=$
$-\alpha_a^2\mu_0^2/(n_r+l+1)^2$ or $p_n=i\alpha_a \mu_0/(n_r+l+1)$.

To find gluonium energy eigenvalues we use the same approach as in 
Refs. \cite{SeA,SeZ}, i.e., we derive an interpolating mass formula for 
$E^2$, which satisfies both of the above constraints. A justification 
is that we do not know the exact form of the QCD potential in the 
intermediate region and exact expression for $E^2$.

Two exact analytic formulae (\ref{Elin}) and (\ref{ECoS}) 
correspond to the linear and Coulomb terms of the potential 
(\ref{Vadj}), respectively. The formulae describe interaction of two 
gluons at large and small distances. Two formulae (\ref{Elin}) and 
(\ref{ECoS}) represent two asymptotes of the exact analytical 
formula for $E^2$ which is unknown. To derive an interpolating 
mass formula, the two-point Pad\'e approximant \cite{Bak} can 
be used,
\be
[K/N]_f(z) =\frac{\sum_{i=0}^Ka_iz^i}{\sum_{j=0}^Nb_jz^j}, 
\label{Pappr}
\ee
with $K=3$ and $N=2$. This results in the interpolating mass formula
\cite{SeZ},
\be
E_n^2 =8\sigma_a\left(2n_r+ l +\frac 32 -\alpha_a\right)
-\frac{4\alpha_a^2\mu_0^2}{(n_r+l+1)^2} +4\mu_0^2. \label{Einter}
\ee
Expression  (\ref{Einter}) is an Ansatz [as the potential (\ref{Vadj})] 
containing the appropriate asymptotic limits, i.e. the two exact 
asymptotic formulae (\ref{Elin}) and (\ref{ECoS}). It allows us to 
reproduce the pomeron trajectory in the whole region.

\section{The Pomeron trajectory}

Regge trajectories being the objects connecting bound state and the 
scattering regions. Expression (\ref{Einter}) has been derived in the 
bound state region. It can be used to derive the Pomeron trajectory in 
the whole region.

The invariant gluonium mass $E^2$ in the bound state region turns into 
the invariant variable $t$ (transfer momentum) in the scattering 
region, i.e. $-E^2=t$. Let us transform Eq. (\ref{Einter}) into the 
cubic equation for the angular momentum $J$, 
\be
J^3 +c_1(t)J^2 +c_2(t)J +c_3(t)=0, \label{JE3}
\ee
where $c_1(t)=2\tilde n +\lambda(t)$,
$c_2(t)={\tilde n}^2 +2\tilde n\lambda(t)$,
$c_3(t)={\tilde n}^2\lambda(t) -\alpha_a^2\mu_0^2/2\sigma_a$,
$\tilde n=n_r+1$, $\lambda(t)=2\tilde n -1/2 -\alpha_a
+(4\mu_0^2 -t)/8\sigma_a$, and $t=E^2$. Equation (\ref{JE3}) has three
(complex in general case) roots, $J_1(t)$, $J_2(t)$, and $J_3(t)$.
The real part of the first root, ${\rm Re}J_1(t)$, gives
the Pomeron trajectory $\alpha(t)$,
\be
\alpha_P(t)=\left\{
\begin{array}{lc} 
\sqrt[3]{-q(t) +\sqrt{Q(t)}} +\sqrt[3]{-q(t) -\sqrt{Q(t)}}
-\frac 13c_1(t), & Q(t)\ge 0; \\
2\sqrt{-p(t)}\cos\left[\frac 13\beta(t)\right]
-\frac 13c_1(t), & Q(t)<0, \label{alt}
\end{array}
\right.
\ee
where $$Q(t)=p^3(t)+q^2(t),\ \ \ p(t)=-\frac 19c_1^2(t)+\frac 13c_2(t),
$$
$$q(t)=\frac 1{27}c_1^3(t)-\frac 16c_1(t)c_2(t) +\frac 12c_3(t),$$
$$\beta(t)=\arccos\left[-q(t)/\sqrt{-p^3(t)}\right].$$

Expression (\ref{alt}) supports existing experimental data and
reproduces the ``soft'' Pomeron trajectory in the whole region of
$t$ (see below); the corresponding parameters $\alpha_a$,
$\sigma_a$ and $\mu_0$ are listened in Table 1. We calculate the
Pomeron trajectory for three different sets of parameters
(methods): I) the typical for light mesons parameter values
$\alpha_s=0.816$, string tension $\sigma=0.15$\,GeV$^2$, and quark
mass $m_q=0.330$ GeV and calculate (according to Ref. \cite{Sim})
the gluonium parameters, $\alpha_a=3\alpha_s=2.448$,
$\sigma_a=\frac 94\sigma=0.338$\,GeV$^2$, and gluon mass
$\mu_0=1.5 m_q=0.495$ GeV (see also Ref. \cite{Halz} for the gluon
mass); II) the parameters $\alpha_a$, $\sigma_a$, and $\mu_0$ are
found from free fit of ZEUS data for the Pomeron trajectory
\cite{ZEUS}; III) include into the fit a $2^{++}$ glueball
candidate at $M=1.710$ GeV \cite{Kirk} by supposing that the
glueball trajectory is the ``soft'' Pomeron trajectory.

\begin{center}
{Table 1: Glueball masses and the Pomeron parameters}

\begin{tabular}{cllll}
\hline \hline
$Method$ & $J$ & $\ \ E_n^{Gl}$ & $ Glueball\ parameters$ &
$\alpha_P(t)~and~\alpha_P^\prime(t)$ at $t=0$ \\
\hline\hline
I & $ 2$ & $\ \ 1.740$ & $\alpha_a=2.448$ & $\alpha_P(0)=1.085$ \\
~&$ 3$ & $\ \ 2.452$ & $\sigma_a=0.338$\,GeV$^2$ &
 $ \alpha_P^\prime(0)=0.250$ GeV$^{-2}$  \\
~&$ 4$ & $\ \ 2.974$ & $\mu_0=0.495$\,GeV & ~ \\
~&$ 5$ & $\ \ 3.408$ & ~ & ~ \\
~&$ 6$ & $\ \ 3.789$ & ~ & ~ \\
\hline
II & $ 2$ & $\ \ 1.984$ & $\alpha_a=2.276\pm 0.041$ &
$\alpha_P(0)=1.084$ \\
~&$ 3$ & $\ \ 2.689$ & $\sigma_a=0.294\pm 0.003$\,GeV$^2$ &
 $ \alpha_P^\prime(0)=0.151$ GeV$^{-2}$  \\
~&$ 4$ & $\ \ 3.164$ & $\mu_0=0.968\pm 0.147$\,GeV & ~ \\
~&$ 5$ & $\ \ 3.549$ & ~ & ~ \\
~&$ 6$ & $\ \ 3.884$ & ~ & ~ \\
\hline
III & $ 2$ & $\ \ 1.695$ & $\alpha_a=2.442\pm 0.044$ &
$\alpha_P(0)=1.113$ \\
~&$ 3$ & $\ \ 2.393$ & $\sigma_a=0.323\pm 0.071$\,GeV$^2$ &
 $ \alpha_P^\prime(0)=0.265$ GeV$^{-2}$  \\
~&$ 4$ & $\ \ 3.904$ & $\mu_0=0.478\pm 0.084$\,GeV & ~ \\
~&$ 5$ & $\ \ 3.330$ & ~ & ~ \\
~&$ 6$ & $\ \ 3.703$ & ~ & ~ \\
\hline\hline
\end{tabular}
\end{center}
From this table, we see that the methods I and III reproduce the
trajectory with the properties of the classic ``soft'' Pomeron. The
intercept and slope estimated by these three methods are:
$\alpha_P(0)=1.09\pm 0.02$ and slope $\alpha_P^\prime(0)=0.22\pm
0.03$ GeV$^{-2}$. Parameters, $\alpha_a$, $\sigma_a$, and gluon mass
$\mu_0$ is close to those predicted by different authors
\cite{Halz,Si}. The corresponding mass of the $2^{++}$ glueball
candidate is around $1.81$ GeV. Masses of gluonium leading states,
$E_n^{Gl}$, have been calculated with the help of the interpolating
mass formula (\ref{Einter}).

\begin{center}
\begin{figure}[ph]
\includegraphics[width=0.9\textwidth]{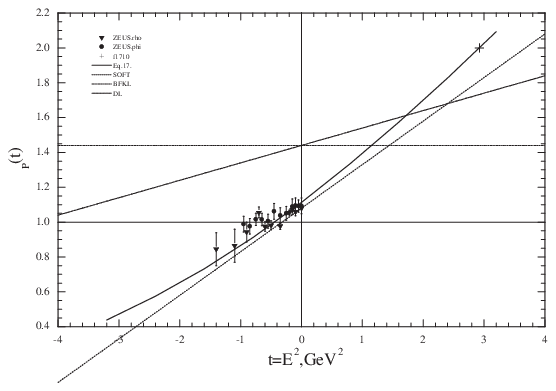}
\caption{\label{fig:PomRT.eps} The Pomeron trajectory. Solid 
curve is the trajectory (\protect\ref{alt}) with the parameters 
found from the fit of combined ZEUS $\rho$ (triangles) and $\phi$ 
(circles) data \protect\cite{ZEUS}, and $2^{++}$ glueball candidate
$f_0(1710)$ (cross) \protect\cite{Kirk}. Other lines show the
classic ``soft'', BFKL, and Donnachie-Landshoff ``hard''
Pomerons.} \end{figure} \end{center}

In Fig. \ref{fig:PomRT.eps} we show the Pomeron with parameters of the 
Method III. The trajectory is linear at $t\ra\infty$ with the 
slope $\alpha_P^\prime =1/8\sigma_a$\,GeV$^{-2}$. In the scattering 
region, the trajectory flattens off at $-1$ for $t\ra -\infty$. 
The first derivative, $\alpha_P^\prime(t)$, is positive in the whole
region, $-\infty <t<\infty$. We see that the experimental data and
simple calculations in the framework of the potential approach
support the conception of the ``soft'' supercritical Pomeron as 
observed at presently available energies.

We can also obtain asymptote for the BFKL-Pomeron
trajectory predicted by pQCD \cite{Lip}. If we take into account 
gluons' spins with the total spin of two interacting gluons $S=2$, 
then the formula \cite{Basd}
\be 
\alpha(E^2)=l(E^2) +S\label{alS}
\ee
gives for the leading ``hard'' pomeron trajectory at large spacelike
$t$,
\be
\alpha_P(t) = 1 +\frac{\tilde\alpha(\vert t\vert)}{\sqrt{1 -
\frac t{4m^2}}}, ~~~ t\rightarrow-\infty.  \label{PSas}
\ee
Pomeron with such properties has been also used to describe the ZEUS
data on the charm structure function $F_2^c$ \cite{ZEUSC}. It was
shown that the two-Pomeron picture (``soft'' plus ``hard'' Pomeron) 
gives a very good fit to the total cross section for elastic $J/\Psi$
photoproduction and the charm structure function $F_2^c$ over the
whole range of $Q^2=-t$ \cite{CDL}. The results of these experiments
and the found higher order corrections \cite{KwMo} make it quite
unclear what a ``hard'' Pomeron is.

There is another explanation of the small $x$ charm production data
at HERA. In many Regge models (see, for instance, Refs.
\cite{Kaid,SeNP}), one-Pomeron exchange gives only dominant
contribution into the cross section. With energy growth, multiple ``soft''
Pomeron (MSP) exchanges and sea quark contributions become important;
these contributions are important just at small $x$. Combined with
the eikonal model the MSP exchanges give the correct energy
dependence of total and total inelastic cross sections \cite{Kaid}
and allow to describe hard distributions of secondary hadrons
\cite{SeNP}. From this point of view, the required ``hard'' Pomeron
discussed in Ref. \cite{DL} effectively accounts for the MSP exchange
contributions.

\section{Conclusion}

After a long period the dominant point of view that Pomeron is a
t-channel process, not a particle, researchers took more seriously
a picture where Pomeron is dual to glueball (gluonium) states,
or even it is a bound state in s-channel. From one side, these 
considerations were encouraged by new data for hard diffraction and 
discovering of Pomeron parton structure. From the other, it was 
caused by a significant theoretical progress with duality concept.

Our previous results obtained for Reggeon trajectories in the framework
of the quark potential model \cite{SeA,SeZ} are in agreement with the
existing experimental data and [for appropriate definition of the
Regge trajectory (see Eq. (\ref{alS}))] with the pQCD predictions on
asymptotic behavior of the trajectories. In this work, an
endeavor to investigate the properties of the Pomeron trajectory in
the framework of the same potential approach has been undertook. This
approach assumes a unify consideration of both the scattering problem
and bound state problem on the basis of solution of the wave equation
for the QCD motivated potential.

It is known, that  using the-fixed-number of particles with a potential 
description can not be used for strict relativistic description.
Strict description of Pomeron presuppose multiparticle description
of the system. For perturbative regime with the Pomeron scattering, the 
dominant contribution comes from BFKL Pomeron. In this work, we have 
constructed Pomeron as a system of two relativistic massive gluons 
interacting by Cornell potential and obtained the interpolating mass 
formula for the squared mass. 

The trajectory obtained is linear at large timelike $t$ and flattens 
off at $-1$ in the scattering region at large $-t$. To reproduce the 
Pomeron trajectory in the intermediate region, we have used the 
interpolating mass formula (\ref{Elin}) for the squared energy 
eigenvalues, $E_n^2=E^2(l,n_r)$, of the two-gluon system. This can be 
justified because we do not know the exact form of the potential in 
this region. The analytic dependence $E^2(l,n_r)$ has
allowed us to reproduce the Pomeron trajectory and calculate its
intercept and slope from the fit of recent HERA data on
$\alpha_P(t)$. These parameters are in agreement with ones obtained
earlier by Landshoff and Nachtmann for the ``soft'' Pomeron.

Perturbative QCD predicts \cite{Bro} different asymptotic behavior of
the Regge trajectories which contradicts to experimental data. The
resolution of this contradiction for the $\rho$ trajectory was
proposed in Ref. \cite{Bro}, namely that the hard QCD part of the
trajectory is weakly coupled and that its contribution will be hidden
until much high energy. However, given the much higher energies at
which the Regge trajectories are known, this argument does not appear
to be of much help and the contradiction remains. A more realistic
explanation can be connected with nonperturbative nature of hadronic
interactions.

Dealing with Regge trajectories, one needs to mention the Regge cuts.
From the Regge viewpoint cuts are expected to become important at
large $-t$. As shown in Ref. \cite{CoKe}, only in a rather limited
intermediate angular region the Regge cuts may be important, $1<-t<3$
\,GeV$^2$ at low energies, $s<60$\,GeV$^2$. At ISR energies the two
Pomeron cut appears to control the whole measured high $-t$ region,
$1.4<-t<14$\,GeV$^2$. However, at larger $-t$ the curvature of the
trajectories enable the poles again to become dominant. The
factorization property of the Pomeron coupling has also been tested
in fixed target experiments \cite{Coo} and found to hold. If the
Pomeron was a cut it would not necessarily lead to factorazable
coupling. The experimental data indicate that it is more likely that
the Pomeron is a pole rather then cut. Therefore, the existing data
and analysis performed in this work confirm the existence of the
Pomeron whose trajectory is nonlinear and coincide with the classic
``soft'' Pomeron at small spacelike $t$.

\section*{Acknowledgments}

The author thanks Professor Uday P. Sukhatme for the kind invitation
to visit the University of Illinois at Chicago where a part of the
work was done and Professor A. A. Bogush for support and constant
interest to this work. The work was supported in part by the
Belarusian Fund for Fundamental Researches.

\end{document}